# VIRTUAL MACHINES AND NETWORKS – INSTALLATION, PERFORMANCE, STUDY, ADVANTAGES AND VIRTUALIZATION OPTIONS

Ishtiaq Ali[1] and Natarajan Meghanathan[2]

[1, 2]Jackson State University, 1400 Lynch St, Jackson, MS, USA

[1]ishtiaq.ali@students.jsums.edu , [2]natarajan.meghanathan@jsums.edu

## ABSTRACT

*The interest in virtualization has been growing rapidly in the IT industry because of inherent benefits like better resource utilization and ease of system manageability. The experimentation and use of virtualization as well as the simultaneous deployment of virtual software are increasingly getting popular and in use by educational institutions for research and teaching. This paper stresses on the potential advantages associated with virtualization and the use of virtual machines for scenarios, which cannot be easily implemented and/or studied in a traditional academic network environment, but need to be explored and experimented by students to meet the raising needs and knowledge-base demanded by the IT industry. In this context, we discuss various aspects of virtualization – starting from the working principle of virtual machines, installation procedure for a virtual guest operating system on a physical host operating system, virtualization options and a performance study measuring the throughput obtained on a network of virtual machines and physical host machines. In addition, the paper extensively evaluates the use of virtual machines and virtual networks in an academic environment and also specifically discusses sample projects on network security, which may not be feasible enough to be conducted in a physical network of personal computers; but could be conducted only using virtual machines.*

## 1. INTRODUCTION

The concept of virtual machines was first developed by IBM in the 1960s to provide concurrent, interactive access to a mainframe computer. Each virtual machine is a replica of the underlying physical machine and users are given the illusion of running directly on the physical machine. Virtual machines also provide benefits like isolation, resource sharing, and the ability to run multiple flavors and configurations of operating systems with different set of software technology and configuration.

Virtualization tools are the main subject of this study therefore; it is important to make a brief description of the available ones in the market. In this study, we have just focused on the VMware products [1] i.e. VMware Workstation and VMware Vcenter Converter. These are open source tools that are run under open source operating systems (OS), with the exception of VMware Server (currently free, but not open source), because of its widespread capabilities running both Windows and Linux platforms as compared to Microsoft Virtual PC or any other virtualization tool, which are only limited to their own software categories. It is important to remark that the similarity level between the virtual and real environment also depends on the virtualization technique [2]. Although the industry uses diverse terms to describe these techniques, they are usually known as emulation, complete virtualization, para virtualization, and operating system (OS)-level virtualization. Some of the widely used are the following virtualization tools:

1. VNUML (Virtual Network User Mode Linux) [3] is an open-source general purpose virtualization tool – enables multiple virtual Linux systems (known as guests) to be run as applications within a normal Linux system (known as the host). As each guest is just a normal application running as a process in user space, this approach provides the user with a way of running multiple virtual Linux machines on a single piece of hardware, offering excellent security and safety without affecting the host environment's configuration or stability.

2. VMware Server [1], as we mentioned previously, is a free virtualization product for Windows and Linux operating systems that implements full virtualization. It allows a physical computer to host some virtual machines, with different guest operating systems.

3. Virtual Box [4] is a x86 virtualization software to deploy virtual machines, destined to desktop computers and enterprise servers, which also implement full virtualization. It allows executing an OS without modification.

4. Qemu [5] is an open source generic emulator that reaches an acceptable emulation speed using dynamic translation. It executes virtual machines under Linux or Windows. It has several very useful commands to manage virtual machines.

5. Xen [6] is an open source virtualization tool, based on the para virtualization technique [7].

The rest of the paper is organized as follows: Section 2 describes the working principle behind virtual machines. Section 3 narrates the installation procedure for creating virtual machines on a VMware workstation. Section 4 describes a performance study experiment conducted on a network of virtual machines and physical host machines and discusses the results obtained for metrics such as network throughput. Section 5 extensively evaluates the use of virtual machines and virtual networks in an academic environment and also specifically discusses sample projects on network security – not feasible to be conducted in a physical network of personal computers; but could be conducted only using virtual machines. Section 6 discusses the various options available for realizing virtualization and creating virtual networks. Section 7 summarizes the contributions of this paper and Section 8 concludes the paper by highlighting the potential advantages of virtualization, especially for academic institutions.

## 2. HOW VIRTUAL MACHINES WORK

VMware (http://www.vmware.com/) is a virtual-machine platform that makes it possible to run an unmodified operating system as a user-level application. The OS running within VMware can be rebooted, crashed, modified, and reinstalled without affecting the integrity of other applications running on the computer. A virtual-machine monitor is an additional layer of software between the hardware and the operating system that virtualizes all of the hardware resources of the machine. It essentially creates a virtual hardware execution environment called a "virtual machine" (VM). Multiple VMs can be used at the same time, and each VM provides isolation from the real hardware and other activities of the underlying system (Figure 1). Because, it provides the illusion of standard PC (Personal Computer) hardware within a VM, VMware can be used to run multiple unmodified PC operating systems simultaneously on the same machine by running each operating system in its own VM. An OS running as a user-level application on top of VMware is called a "guest OS." The native OS originally running on the real hardware is called the "host OS."

VMware is low-level enough to make a guest OS appear to be receiving hardware interrupts (such as timer interrupts) and behave as if it were the only OS on the machine. At the same time, it provides isolation so that a failure in or misbehaving of a guest OS does not affect other guest OSs or the underlying system. For instance, a guest OS crashing will not crash the underlying system. As opposed to a software simulator, much of the code running in a VM executes directly on the hardware without interpretation. Operating systems currently supported as guest operating systems under VMware include Windows 95/98/2000/NT, FreeBSD, Solaris, Novell Netware, DOS, and Linux, all of which run unmodified. Theoretically, any OS that can run on an x86 architecture can run as a guest OS, since it will see a complete virtualized PC environment. For host operating systems, VMware currently runs and is supported on Windows Vista, XP, 2000/NT and Linux.

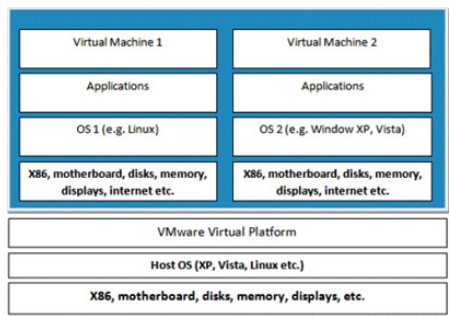

**Figure 1:** Two Virtual Machines Hosted on One Host Operating System

There are number of VMware appliances (guest OS) available at the website: http://www.vmware.com/appliances. Some of them are free and some of them are trial versions. Software appliances are developed by Independent Software Vendors (ISVs) and provided for software applications to be pre-installed and pre-configured. A software appliance generally includes a customized and optimized operating system and the software application packaged within it. A virtual appliance is defined as a minimal virtual machine image that contains the software appliance designed to run in a virtualized environment. But, we can also build customized appliances or packages for teaching, software experimentation as well as performance and network analysis.

In order to build our own customized appliance, we will need to install VMware workstation. The installation of guest operating systems and required applications is similar to the installation of host operating system. The only difference is that the user must first set or allocate the resources from the host OS before installation of the guest OS.

## 3. INSTALLING A GUEST OS USING VMWARE WORKSTATION ON WINDOWS XP PROFESSIONAL OS

In this section, we will demonstrate how to install guest OS using Unbuntu desktop ISO image on Windows XP Professional OS using VMware Workstation 6.5. We have already downloaded and installed the VMware workstation 30 day trial edition for this demonstration. VMware Workstation can manage several guest OS as an application as compared to the VM player and gives the option to customize the guest OS according to the need or required configuration. VMware has a hosted architecture to virtualized I/O that allows it to co-exist with a pre-existing host operating system: VMApp, VMDriver and VMM. VMApp is a simple application that allows users to install other operating systems. This application uses a driver (VMDriver) loaded into a host OS to establish the privileged virtual machine monitor (VMM). This virtual machine runs directly on the hardware.

The installation of VMware Workstation is quite simple; its installation is not different than any other Windows application software. The only difference is that after the installation, the system has to be restarted because the VMware workstation will install the network adapter on the host machine. In our example (refer Figure 2), we have VMware workstation install two network adapters on the host machine: VMware network adapter VMnet 8 and VMware network adapter VMnet1. The VMnet8 adapter is used for the bridge networking configuration for guest OS where as the VMnet1 adapter used for the NAT (Network Address Translation).

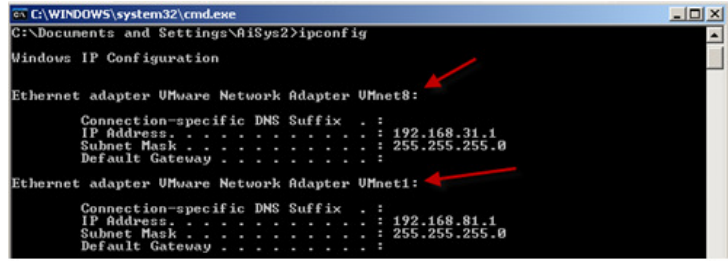

**Figure 2:** VMware Workstation (VMnet 8 and VMnet 1) Installed on Host OS

## 3.1. Installation Steps

In this section, we first describe the four major audio steganography algorithms: Low-bit encoding, Phase encoding, Spread spectrum coding and Echo data hiding. The disadvantages associated with these algorithms can be exploited for steganalysis [16].

Step 1: We have already downloaded the free Ubuntu Desktop version 9.04-i386 (32-bit) ISO image from http://www.ubuntu.com/GetUbuntu/download. But, the ISO image needs to be burned first on the optical disk before the software can be installed. VMware workstation allows a user to install the guest OS using just the disk image. It is also important to download the right image, for example, if the host machine is 32-bit based, we need to use a 32-bit image or vice versa.

Step 2: After loading the VMware Workstation, select "*New Virtual Option*" Option as shown in Figure 3; Select "*Typical*" (Figure 4) and Click "*Next*". This option will allow the user to accept the defaults or specify values for customizing the hardware. Users can select "*Custom*" if they want to build a machine other than what is specified by VMware common guest OS or need to specify the I/O adapter type for SCSI adapters: BusLogic, LSI Logic, or LSI Logic SAS or to specify to create IDE/SCSI virtual disk.

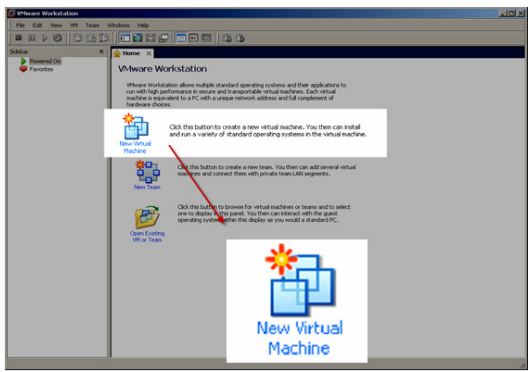

**Figure 3:** Loaded VMware Workstation

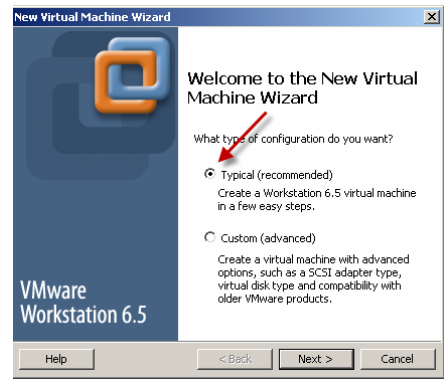

**Figure 4:** Choosing a VM Installation Option

<u>Step 3:</u> The next screen will be prompted with the option of installing the guest OS from the optical disk or from the disk image. We will be using Ubuntu ISO image file, stored on the local machine, as shown in Figure 5. But, if the VMware Workstation can detect the OS in optical disk drive on media, this screen may not appear. As shown in Figure 5, the Ubuntu image has been selected. Clicking "Next" will display the prompt for the name as well as guest OS user (should all be in lower case) and password as shown in Figure 6.

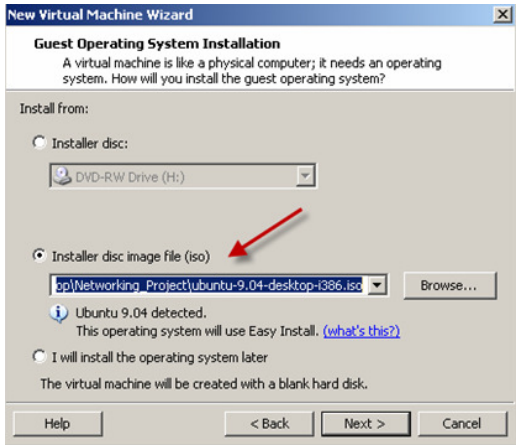 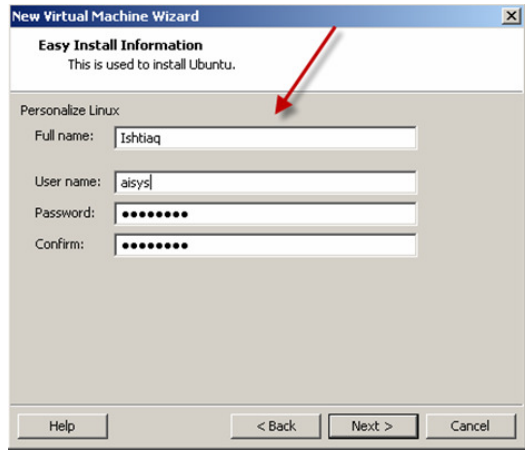

**Figure 5:** Guest OS Source Location          **Figure 6:** Username and Password Screen

<u>Step 4:</u> Type the user name and password. Clicking "*Next*" will prompt for the virtual machine name and the location for the virtual machine installation folder as shown in Figure 7.

<u>Step 5:</u> Type the virtual machine name for example "*Ubuntu*" and location for the virtual machine installation folder as shown in Figure 7. Clicking "Next" will prompt the user to specify the hard disk configuration.

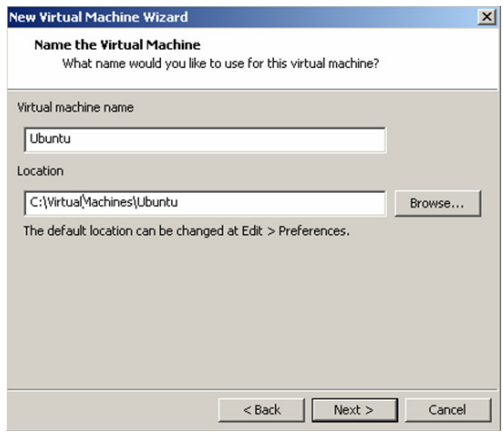 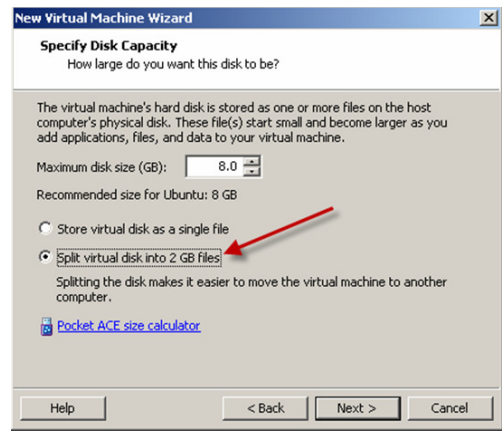

**Figure 7:** Guest OS Name, Installation Location   **Figure 8:** Guest OS Hard Disk Configuration

<u>Step 6:</u> In this step, the user will specify the total disk capacity for the virtual machine. VMware workstation 6.5 provides a new technique for either keeping the guest virtual machine as a single file or splitting the virtual machine into 2 GB segments. This approach is quite beneficial if one wants to write the virtual machine on disk and move it to a different location or computer.

For example, the virtual machine image can be distributed for Computer Networks or Network Security classes with all the required software installed on it. Then, the students can load the image on their machine using the free VM player. For Unbuntu, we have allocated 8 GB disk space, split into segments of 2 GB space, as shown in Figures 8 and 9.

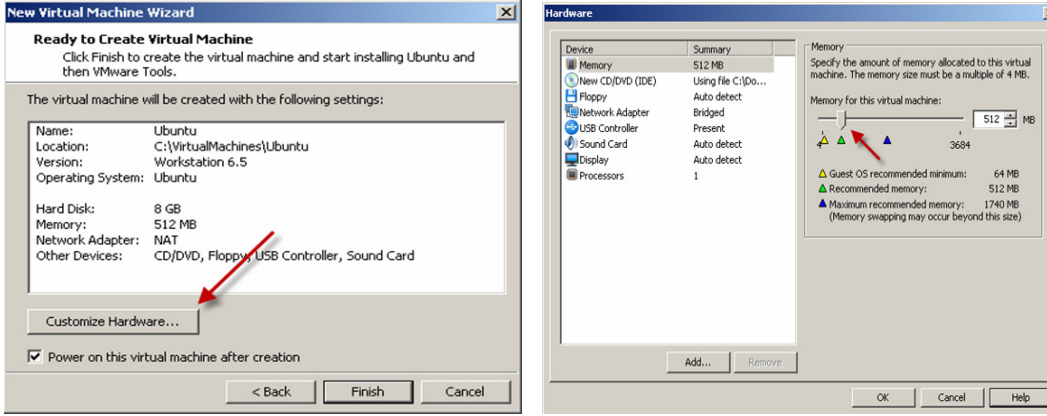

**Figure 9:** Guest OS Hardware Configuration  **Figure 10:** Customization of Guest OS Memory

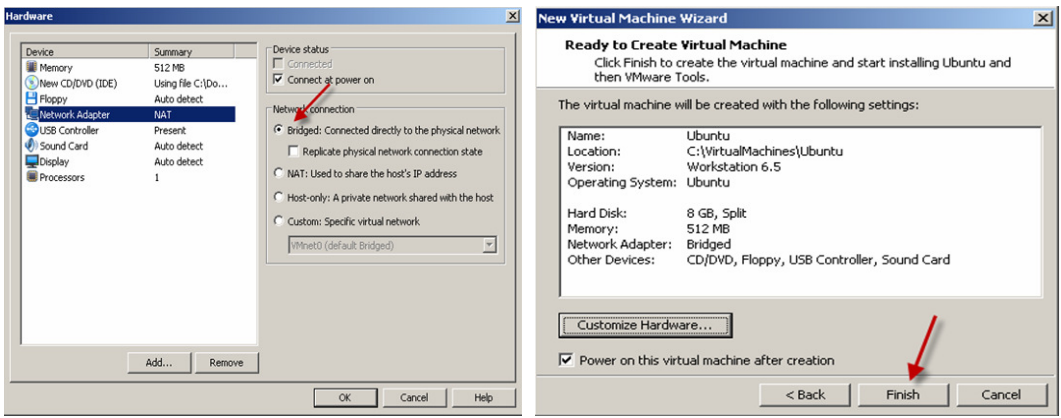

**Figure 11:** Bridge Configuration for Guest OS  **Figure 12:** Guest OS Ready to Install

Step 7: The final step will show the selected configuration for the virtual machine. The user also has the option at this point to make any final changes by clicking on the "Customize Hardware," link as shown in Figures 9 and 10. The VMware workstation provides the user with the option to change the memory for the guest operating system virtual machine (Figure 10).

User can change the network configuration by selecting the "*Network Adapter*" option from the list as shown in Figure 11. In this panel, the user can change their virtual network adapter and add additional virtual network adapters. The virtual machine should be powered off before adding or removing a network adapter. The user can set the following options under the device section:

Connected: Connects or disconnects the virtual network adapter while the virtual machine is running.

Connect at power on: automatically connects the virtual network adapter to the virtual machine when powered. The user can make changes to the following options in the *Network connection* section when the virtual machine is powered on or powered off:

*Bridged:* If the host computer is on an Ethernet network, bridged networking is often the easiest way to give virtual machine access to the network. With bridged networking, the virtual machine appears as an additional computer on the same physical Ethernet network as the host. The virtual machine can then transparently use any of the services available on the network to which it is bridged, including file servers, printers, and gateways. Likewise, any physical host or other virtual machine configured with bridged networking can use resources of that virtual machine.

*Replicate physical network connection state:* This option is very useful, if the user has selected a bridged network and the virtual machine is installed on a laptop or other portable device. This option will automatically renew the IP address of the virtual machine as user moves from one wired or wireless network to other.

*NAT (Network Address Translation):* The NAT option could be selected to connect to the Internet or other TCP/IP network using the host computer's dial-up networking connection if the user cannot or do not want to give the virtual machine an IP address on the external network. A separate private network is set up on the host computer. The virtual machine obtains an address on that network from the VMware virtual DHCP server.

*Host-only:* the virtual machine is connected to the host operating system on a virtual private network, which normally is not visible to the outside host. Multiple virtual machines could be configured with host-only networking on the same host and on the same network.

*Custom:* to set up a more complex networking configuration, the user can customize setup for one or more of the virtual network adapters. After selecting 'Custom,' the user can choose a virtual switch from the drop-down menu. This connects the virtual machine's adapter to that switch. All virtual machines running on the same host computer and connected to the same virtual switch are on the same virtual network.

<u>Step 8:</u> The final step is to click on "Finish" to start the installation of the guest operating system as shown in Figure 12; it will usually take 20 to 30 minutes depending on the host system configuration and speed. Figure 12 shows the summary of our virtual machine's final hardware configuration. Figure 13 shows the Unbuntu guest OS installation in progress and Figure 14 shows the completely installed Ubuntu virtual machine running the Windows XP Professional operating system.

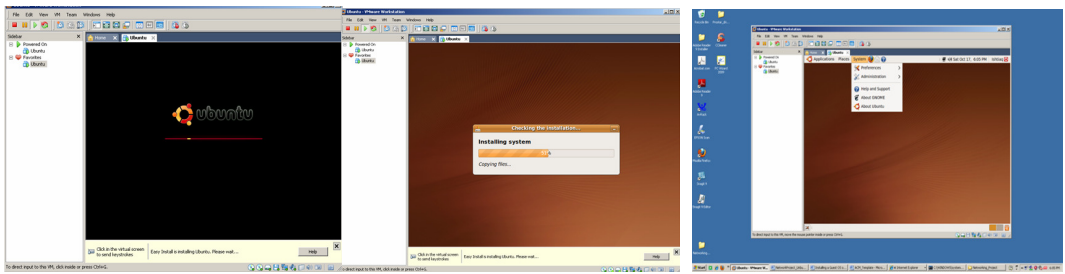

**Figure 13:** Ubuntu Guest OS Installation in Progress   **Figure 14:** Ubuntu Installation Done

## 3.2. Switching between Guest OS and Host OS

It is important to mention how the user can interact with the host and guest OS at the same time. Clicking inside a virtual machine will make the interaction active in the virtual machine. If the user wants to interact with the host OS; the user has to hold the Ctrl+Alt keys to release the mouse to the host OS. To avoid the 'Ctrl+Alt' key combination, the user can install an important piece of software called VMware tools for the VMware virtual machines or guest OS. A VMware tool provides several features like shared folders drag and drop feature between host and guest OS. Other features include time synchronization, automatic grabbing and releasing of the mouse cursor, copying, pasting between guest and host. The installers for the VMware tools for Windows, Linux, and FreeBSD guest OS are built into the VMware workstation as an ISO image file. But in older versions of VMware workstation, the user can install the VMware tool after installing the guest OS using the menu option of VMware Workstation *VM>Install VMware Tools*.

### 3.3. Converting Physical Host Machine to Virtual Machine using VMware VCenter

VMware also provides free VMware Vcenter Converter to convert physically installed operating system into virtual machine. It can convert Microsoft Windows and Linux based physical machines to virtual machines. This free software can be downloaded from http://www.vmware.com/products/converter/ with installation and conversion instructions. In this paper, the host machine Windows XP has been converted into virtual machine using VMware Vcenter Converter for performance testing. The setup of these experimental machines is discussed in Section 4 on the Performance of Virtual Networks.

### 3.4. Licensing of Virtual Machines

Virtualization is the IT industry's fastest growing technology with major cost-reduction benefits, but at the same time, the technology raises licensing issues for vendors and consumers. For example, prior to 2008, Microsoft considered virtual OS as an independent operating system and hence required a separate license. But recently, Microsoft has made significant changes in their software licensing for virtual machines from installation-based to an instance based. The revised policy on licensing for different Microsoft products family can be found at: http://www.microsoft.com/licensing/about-licensing/virtualization.aspx.

## 4. PERFORMANCE OF VIRTUAL NETWORKS

Virtualization does provide an excellent flexibility and portability, but can also introduce degradation in network performance, especially in high performance throughput and low latency devices. This section analyzes the overhead associated with VMware-based virtual networks. The results from our experiments can be used as benchmarks and as reference for comparison testing.

### 4.1. Experimental Setup

The experimental layout of the machines is shown in Figure 15. All the tests have been developed on the following host and guest operating systems:

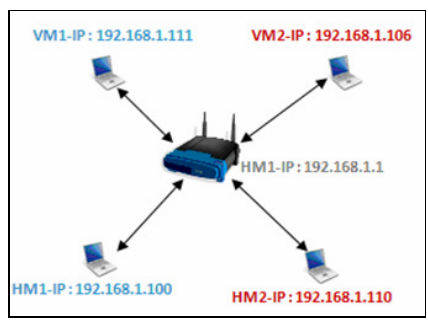

**Figure 15:** Experimental Layout of the Physical Host Machines and Virtual Machines

(i) *Host Machine Configuration – Host Machine* 1 (HM1): Operating System – Microsoft Windows Vista Home Premium. System Manufacturer – Hewlett-Packard Notebook PC, System Type – x64-based PC, Processor, Intel(R) Dual Core(TM); CPU – P7350 @ 2.00GHz, 2000 MHz, 2 Cores, 2 Logical Processors; Installed Physical Memory 6 GB; Connected to Ethernet via wireless router.

(ii) *Virtual Machine on* HM1 (VM1): Same hardware devices as HM1; Allocated memory is 1 GB, allocated hard disk is 127 GB; Networked using VMware workstation Bridge configuration, which is directly connected to the local network with its own IP address. On this virtual machine, we installed Windows Server 2003 Service Pack 3.

(iii) *Host Machine Configuration – Host Machine* 2 (HM2): Operating System – Microsoft Windows XP Professional; System Manufacturer – ProStar Notebook PC; x32-based PC, Processor Intel Pentium 4, 3.2 GHz Processor; Installed Physical Memory (RAM) – 3 GB; Connected to Ethernet via wireless router.

(iv) *Virtual Machine on HM2 (VM2)*: Same hardware devices as HM2; Allocated memory 504MB, allocated hard disk 5 GB; Networked using VMware workstation Bridge configuration, which is directly connected to the local network with its own IP address. The HM2 machine has been cloned to a virtual machine using VMware Vcenter Converter software.

## 4.2. Netperf Software

The above experimental setup was used to determine the maximum throughput of the virtual machine network in client/server environment using the Netperf tool [8]. Netperf makes measurements at the transport layer of the OSI model. Its primary focus is to measure the throughput and client/server interface performance using either Transmission Control Protocol (TCP) or User Datagram Protocol (UDP). Since Netperf is a client-server application, it has two executables - Netperf and Netserver. Netserver must be executed in order for Netperf to connect and retrieve the appropriate results. Netperf runs on the client and Netserver runs on the server. When Netperf is invoked on the client system, a control connection to the remote (Netserver) gets established. This connection is used to pass test configuration information and test results to and from the remote system. Once the control configuration is established and the configuration information has been passed, another connection is initiated for retrieving the measurements according to the control configuration.

**Netperf Usage**
The usage for Netperf server and client is as follows:

*Netperf Server Usage*

netserver [options]
*-h to display the help or usage of the server*
*-p port number to specify the port for the server default port is 12865*

*Netperf Client Usage*
The Netperf client command-line options are divided into two categories: (a) Global and (b) Test-specific command-line options. Both category commands can be provided with single command line separated with double dash
*netperf [global options] -- [test-specific options]*

### 4.3. Bulk Virtual Machine Traffic Measurement

This section describes the Netperf test that determines the performance of bulk data transfers. This type of network traffic is common in many network transactions, from File Transfer Protocols (FTPs) to accessing data on shared network drives. In the following tests only the VM1 machine network performance will be tested for the fact that this virtual machine has Windows Server 2003 installed. The Netperf client program will be executed from HM2 and VM2 i.e. HM2 to VM1 and VM2 to VM1.

*TCP_Stream:* This test sends bulk TCP data packets to the Netserver host, and determines the throughput that occurs during the data transfer.

***Netperf command***
*netperf -H  [IP running Server]  -l  60*
*netperf -H  [IP running Server]  -- -m 2048*

whereas -l option is used to set the test duration for 60 and 10 seconds (the default is 10 seconds).

In this experiment, the Netperf server automatically sets the message size to the size of the socket send buffer on the local system. The average throughput (Table 1) from the physical host to virtual host is 21.73 Mbps as compared to physical host to physical host machine 22.39 Mbps, ignoring the send message size from less than 1KB to 2KB message. This is not very significant difference in the throughput, which means that the virtual machine can perform same as the physical machine. Although the experiment is conducted with a very small local network, it shows that running a virtual machine on a physical machine does not drastically affect the host or virtual machine network performance.

*TCP_RR*: This experiment tests the performance of multiple TCP request and response packets within single TCP connection. This simulates the procedure that many database programs use – establishing a single TCP connection and transferring database transactions across the network on the connection. The following Netperf client command is used:

*netperf -t TCP_RR -H [Server Address] -l 60*
*netperf -t TCP_RR -H [Server Address] -l 60 -- -r 32, 1034*

In TCP Test Response Request, -r option sets the size of the request or response message or both. -r 32, 1024 sets the size of the request message to 32 bytes, and the response message size to 1024 bytes. The average transaction rate shows that (Table 2) 604 transactions were processed per second with message size for both the request and response packets was set to 1 byte in default test. Then to get some realistic situation, we set the request message to 32 bytes, and the response message size to 1024 bytes. Even with the larger message size, the transaction rate did not drop dramatically from host physical machine to virtual machine.

**Table 1:** TCP_Stream Request/ Responses

| Trans. Path | Recv. Socket Size bytes | Send Socket Size bytes | Send Messag Size bytes | Elapsed Time Secs. | Throughput 10^6 bits/sec |
|---|---|---|---|---|---|
| HM2-VM1 | 8192 | 8192 | 8192 | 60 | 24.14 |
| VM2-VM1 | 8192 | 8192 | 8192 | 60 | 22.27 |
| HM2-VM1 | 8192 | 8192 | 2048 | 10 | 19.32 |
| VM2-VM1 | 8192 | 8192 | 2048 | 10 | 19.82 |
| HM2-HM1 | 8192 | 8192 | 8192 | 60 | 24.14 |
| HM2-HM1 | 8192 | 8192 | 2048 | 10 | 20.64 |

**Table 2:** TCP Request/ Response Test for VM1 (192.168.1.111) – Windows Server 2003

| Trans. Path | Socket send bytes | Size Recv. Bytes | Req. size bytes | Resp. size bytes | Elapsed Time secs | Trans rate/sec |
|---|---|---|---|---|---|---|
| HM2-VM1 | 8192 | 8192 | 1 | 1 | 60 | 604.43 |
| HM2-VM1 | 8192 | 8192 | 32 | 1034 | 60 | 492 |
| HM2-HM1 | 8192 | 8192 | 1 | 1 | 60 | 596 |
| HM2-HM1 | 8192 | 8192 | 32 | 1034 | 60 | 511.4 |

# 5. USE OF VIRTUAL NETWORKS IN ACADEMIC ENVIRONMENT

Many universities typically provide an account for students, often with limited access and privileges, in their servers dedicated for a particular systems course or a programming course. But, it is often difficult to expect universities creating more than one account per student. If students have to run multiple processes (e.g., a multi-user chatting application), they would have to typically open multiple terminals within the same account and run the processes at different port numbers. Even in universities with dedicated labs for the courses, students rarely get chance to simultaneously run their processes on multiple physical machines and observe the interaction between these processes [19]. For such scenarios, students could download pre-built Linux-based appliances (without any restriction on licensing as well as relatively lower resource overhead than Windows-based appliances) using which they can simultaneously run several virtual machines and test their applications. Virtual machines play a significant role in reducing the need for several physical host machines to run multiple processes.

In addition, if students are interested in trying out certain special software for their course or research projects, they would have to go through the instructors/ universities for obtaining permissions as well as requiring the institution to install the software. Virtual machines can reduce the administrative overhead for the Information Technology (IT) divisions in an institution and also simultaneously enhance student creativity and performance. With virtual machines, students have several options to try out. They could download pre-built virtual appliances (some may be completely free and others may be available in trial versions) and install. Students can further install any required programming language compiler, software development kit on a virtual machine without affecting their personal machine (i.e. the host). After downloading and installing the virtual machine they can connect their virtual machine to their home based router either using VM player Bridge adapter, which will probably be the best option for the fact that the virtual machine will have its own IP address similar to the host machine. The other option is to use NAT (Network Address Translation) adapter to connect to the router indirectly via the host machine. After all, a virtual machine breakdown will neither affect the physical host machines nor the network.

A virtual machine is the best candidate for courses related to Network Security. In order for students to run vulnerability related programs against the machines, they would have to first have a machine on which they can create such security risks and then create their programs or run commercially available programs to detect and/or study different types of attacks on a machine. Most of the network security related projects are best suited for Linux-based virtual machines. Again, the university level account will not be the best option for such projects due to the fact that students will need more privileges on their account for administration purposes as well as to create different privilege levels for the account as per the needs of the experiments. The advantage of running such exercises on a virtual network is that none of the damaging or questionable traffic can get generated on any of the production network, and all of the project could be run not just from the lab but from a properly configured remote location. VMware machines allow for the creation of simple files or group of files that can be distributed with the entire configuration necessary to demonstrate topics in a way that does not negatively impact the device or the network the device is running on.

Virtual machines could be widely adopted in academics (for example, in many courses), because the main objective of virtualization is to reduce the cost, and keep the host system unmodified and make the host portable and manageable as much possible. Students will have an accessible environment to work on their projects both from on campus and remotely. A very feasible and cost-effective solution is possible that closely resembles real-life environment, easily adaptable to the changing needs of the courses without the overhead of IT resources and cost. Several options can be considered to provide such facility to students. Below, we explore the use of virtual machines for some of the commonly studied problems in computer and network security related courses.

## 5.1. Stack-based Buffer Overflows

Buffer overflow attacks have been around for quite some time and they will still be a problem to be explored in the near future. On many C implementations, it is possible to corrupt the execution stack by writing past the end of an array declared 'auto' in a routine [9]. Most Linux machines do provide kernel-based stack overflow protection e.g. kernels randomize stack addresses to make it difficult to predict locations of shell code. Linux-based scripts can be used to turn the stack protection on or off and such scripts must run under root privileges which is again not possible with university provided account; neither will be the best interest for university to run such experimentations.

## 5.2. Ping Tracing through Firewalls

This project will demonstrate how a firewall filters incoming traffic. For this project, the guest VM will need two pieces of software: a packet capture tool and a SSH client. SSH Secure Shell [10] or PuTTY [11] is a freely available client that will work well, and Wireshark [12] is similarly available for packet capturing. This captures traffic both on the external interface of the firewall, and the interface of the guest VM (connected to the internal interface of the firewall).

## 5.3. Port Scanning and Advanced Probes

A useful Linux distribution with plenty of security-related tools is Knoppix-STD [13]. This Linux live-CD can be used in standard computer labs by booting to the CD, but any commands run will impact the network directly. Instead, a virtual network can be quickly setup to probe specific virtual machines and identify weaknesses in their configuration.

## 5.4. Using *Nmap* and *Nessus*

Nmap ("Network Mapper") is a free and open source utility for network exploration or security auditing [14]. The Nessus vulnerability scanner [15] is the world-leader in active scanners, featuring high-speed discovery, configuration auditing, asset profiling, sensitive data discovery and vulnerability analysis. Both tools are commonly used on Linux-based platforms. Students can use the nmap scan tool on a specific target machine for active ports from a root shell with the defaults and the tool will return open scanned ports for the targeted virtual machine. This will help students to identify any weakness in the configured security of the target machine. Nessus can also be used on virtual machines to build the complete vulnerability report. In fact, Nessus does provides a VMware virtual image to its subscribers as a professionally built appliance and also available on a variety of hardware appliances.

## 5.5. Network Intrusion Detection

Students can use a Network Intrusion Detection System (NIDS) to detect attacks through the network. The popular tool of for detection is Snort, an open source signature-based NIDS [16]. Students can download non-subscription based rules set from the http://www.snort.org website. Snort will also require root privileges in order to copy the rule sets to "*/etc/snort*" folder, setting up RULE_PATH. It is required to run Snort as root instead of regular user.

## 5.6. Man-in-the-Middle Attack

Man-in-the-Middle attacks can be launched on the Address Resolution Protocol (ARP), a relatively simple link-layer protocol to retrieve the hardware address of a machine using its IP address. We explain the ARP protocol and the Man-in-the-Middle attack on ARP through the following example: When a host A wants to send a packet to an IP address on the same LAN, it sends an Ethernet broadcast requesting the MAC address of a node with a particular IP address. When a targeted host B sees a request for its IP address, it will send a reply with its MAC address. Host A will then cache the result for a short period of time, using that MAC address for future packets sent to host B. However, there is no built-in form of authentication in ARP; therefore, replies can be easily spoofed. The Ethercap suite [17] can be used for launching ARP-based Man-in-the-Middle attacks on a LAN. It features sniffing of live connections, content filtering on the fly and many other interesting tricks. In this experiment, a student will require at least 3 machines. Students can download Linux-based virtual appliances and create copy of the Linux virtual machines. Now with two Linux virtual machines and one host windows machine, student can create a setup for this experiment. Student can then send ICMP echo message from one of the machine to both of the target machines.

## 6. VIRTUALIZATION OPTIONS IN ACADEMIC ENVIRONMENT

We evaluate the following different options for setting up a lab (network) of virtual machines in an academic campus environment.

### 6.1. VMware Workstation

The main software needed would be "VMware workstation" by VMware. VMware workstation is powerful desktop virtualization software that allows users to run multiple x86-based operating systems like Windows, Linux and Netware and their applications simultaneously in fully networked portable virtual machines. The advantage of this option as compared to the traditional option is that there is no need for additional space to host the hardware, the software could run on the current PCs in the classroom, and students will have their own portable virtual machine which will meet or exceed their need. Student will have the option to load either Windows or Linux based operating system for their project needs. The disadvantage to this option is the initial cost of purchasing the software, and also it would be hard to setup and administer the individual virtual machines.

### 6.2. Microsoft Virtual PC

Microsoft Virtual PC [18] can be an option, but this virtual machine can only support Microsoft OS and does not support open source or other vendor operating systems, for example Linux, Mac OS etc. Virtual PC may be attractive to those schools with a Microsoft software licensing agreement, as it is designed to work with Windows servers, but it has significant limitations, especially for network use.

### 6.3. QEMU and Xen Options

QEMU [5] and Xen [6] have more usable features than Microsoft products, but are only supported on Linux host computers, and are more difficult to configure and install. Once installed, the virtual machines actually have more flexibility with network configurations, as an unlimited number of virtual networks can be configured. The main issue with using these products is their lack of support for Microsoft guest operating systems. Windows is unsupported (though has reportedly worked) on QEMU, and will not be supported on Xen, until the release of a new processor virtualization technology from Intel and AMD.

### 6.4. VMware GSX Server

VMware GSX server [1] can be the best candidate for an academic setting as it is enterprise-level virtual infrastructure software for x86-based machines/servers. VMware GSX Server allows virtual machines to be remotely managed, automatically provisioned, and standardized on a secure, uniform platform. Required operating systems and related applications can reside in multiple virtual machines on a single host physical hardware. VMware GSX server provides broad hardware support by inheriting device support from the host operating system. The product's robust architecture and ability to integrate into Microsoft Windows and Linux host environments make it quick and easy to deploy and manage. VMware GSX server runs as an application on a host operating system; manages and remotely controls multiple servers running in a virtual environment. The advantages of this option are the ability for customization and provisioning of virtual images/configurations, a user interface for easy management of multiple virtual machines sessions, secure access with OpenSSL, secure remote management, automated monitoring and control, web interface for instructors and students to authenticate and access their virtual machines, integration with campus Active Directory and support for popular Linux distributions. The VMware GSX server has high initial cost for the software and hardware and there would be also an on-going maintenance cost. But the benefits and outcomes would be far greater than the traditional option of buying dedicated hardware and software to run only operating system per machine.

## 7. SUMMARY OF CONTRIBUTIONS

This paper contributes to the literature on Virtual Machines and Virtualization in the following aspects: (i) We provide a tutorial-like step-by-step procedure to install a guest operating system on VMworkstation, a commonly used virtual machine environment, which is running on a Windows host operating system; (ii) We describe a small experiment that has been conducted to compare the performance of a network of virtual machines and the performance of a network of physical machines and measured the network throughput obtained for bulk traffic scenarios such as file transfers; (iii) We extensively evaluate the use of virtual machines in an academic environment and discuss the various virtualization options that are currently available; and (iv) We discuss the potential advantages associated with using virtual machines for security-related projects and experiments in a campus setting. In this direction, we provide a sample list of projects on network security, which may not be feasible enough to be conducted in a physical network of personal computers; but could be conducted only using virtual machines.

## 8. CONCLUSIONS

Virtualization can create real world business environment as closely as possible in an academic setting, so that students can interact with technologies just as they would in a work setting. In educational institutions, it is not always possible to provide such laboratory which can provide software as well configuration to each discipline of the institutions; the reality in most institutions is to have shared laboratories, used by different students and disciplines. This problem can be alleviated by the use of virtual machines, allowing each student to build his/her own network experiment, using the appropriate topology, and thus not disturbing the other activities running in the lab [20]. Student(s) who would like to understand for example network protocols or security issues can freely download already pre-built virtual appliances and install the required software to work on their specific projects. The performance study conducted in this paper, although on small scale, shows that there would be no significant performance overhead on a virtual network of host machines and virtual machines. In conclusion, virtualization is a new growing trend in the IT industry. Businesses as well as educational communities can equally be benefited from it despite the overhead involved in setting up a virtual network.

## ACKNOWLEDGMENTS

This research has been partly funded through the U. S. National Science Foundation (NSF) CCLI/TUES grant (DUE-0941959) on "Incorporating Systems Security and Software Security in Senior Projects." The views and conclusions contained in this document are those of the authors and should not be interpreted as necessarily representing the official policies, either expressed or implied, of the funding agency.